\documentclass[twocolumn,times]{aastex62}

\graphicspath{{./}{figures/}}

\received{ \today}
\revised{} 
\accepted{} 
\submitjournal{ApJL}

\shorttitle{Ultra-compact X-ray binary origin Evolution of M71E pulsar binary}
\shortauthors{Yang et al.}

\begin{document}

\title{PSR J1953+1844 probably being the descendant of an Ultracompact X-ray binary}

\correspondingauthor{J.L. Han}
\email{hjl@bao.ac.cn}

\author{Z.~L. Yang}
\author{J.~L. Han} 
\author{W.~C. Jing}
\author{W.~Q. Su}

\affil{National Astronomical Observatories, Chinese Academy of Sciences, Beijing 100101, China}
\affil{School of Astronomy and Space Sciences, University of Chinese Academy of Sciences, Beijing 100049, China}
\affil{CAS Key laboratory of FAST, National Astronomical Observatories, Chinese Academy of Sciences, Beijing 100101, China}

\begin{abstract}
PSR J1953+1844 (i.e., M71E) is a millisecond pulsar (MSP)in a 53 minute binary orbit discovered by the Five-hundred-meter Aperture Spherical radio Telescope. The mass function from pulsar timing is $2.3\times10^{-7}$ $M_\odot$. The possible redback origin of this system has been discussed by Pan et al. We discuss here an alternative evolution track for this binary system, namely that PSR J1953+1844 is a descendant of an ultra-compact X-ray binary (UCXB), which has a hydrogen-poor donor accreting onto a neutron star (NS) with an orbital period of $\leq1$ hr. We noticed that some of UCXB systems hold an accreting millisecond X-ray pulsars (AMXPs) and a donor with a mass of about 0.01 M$_\odot$. M71E has a very similar orbit to those of AMXPs, indicating that it might be evolved from a UCXB similar to PSR J1653--0158. The companion star of M71E should be significantly bloated and it most probably has a carbon and oxygen composition, otherwise a low inclination angle of the orbit is required for a helium companion. The discovery of this M71E binary system may shed light on when and how an NS in a UCXBs turns into a radio pulsar.
\end{abstract}

\section{Introduction}

Soon after the discovery of a radio millisecond pulsar \citep[MSP;][]{1982Natur.300..615B}, it was suggested that such a fast-rotating object may be originated from an ultracompact X-ray binary (UCXB). Nowadays a large number of accreting millisecond pulsars have been found in UCXBs \citep{2010HiA....15..121W,2018ApJ...858L..13S,2021ApJ...908L..15N}. However, the transition from accretion-powered MSPs in UCXBs to these rotation-powered pulsars is not yet well understood \citep{2013ApJ...768..184H}. Traditionally, eclipsing MSPs in a tight orbit (orbital period $P_{\rm orb}\leq$ 1 day) with a companion of mass 0.1--0.4 $M_\odot$ are classified as redbacks, while those MSPs orbiting a companion of mass $\ll0.1~M_\odot$ are classified as black widows \citep{Chen_2013,2013IAUS..291..127R}. The two systems are the subclasses of spider pulsars whose companion stars are suffering from severe mass loss. Among them, black widows with an orbital period less than 2 hr, e.g, PSR J1653$-$0158 \citep{2020ApJ...902L..46N}, PSR J1719$-$1438 \citep{2011Sci...333.1717B} and PSR J0636+5128 \citep{2018ApJ...864...15K}, are believed to be the descendants of UCXBs. It was proposed that the pulsar wind may play an important role in this transition process, and binary pulsars will form  {from UCXBs} with an orbital period of $\sim$1.5 hr \citep{2011Sci...333.1717B,2013ApJ...768..184H}. 

Recently a binary pulsar PSR J1953+1844 (i.e. M71E,  {because it is probably located in the globular cluster M71}) was discovered by the Five-hundred-meter Aperture Spherical radio Telescope \citep[FAST;][]{2006ScChG..49..129N,2011IJMPD..20..989N} during the FAST Galactic Plane Pulsar Snapshot (GPPS) survey \citep{2021RAA....21..107H}. The follow-up timing
observations have revealed that M71E has an orbit period of 53 minutes \citep{53-min}.  {Its short orbit period of 53 minutes breaks the previous record for the orbit period of 75 minutes for PSR J1653--0158 \citep{2020ApJ...902L..46N}; no eclipses and significant orbital period changes have been detected yet \citep{53-min}.} The nature of this system is not well-understood yet. This system was interpreted as an intermediate between redbacks and black widows by \citet{53-min}.  Based on a low mass function of $2.3\times10^{-7}~M_\odot$ and the strikingly short orbital period in the range of UCXB orbital periods, here we discuss an alternative evolution scenario for this binary system that is a newly evolved descendant of a UCXB, as a complementary theory to \citet{53-min}. We discuss in this Letter the evolution and fate of this binary system, and suggest that the companion of M71E 
is a hydrogen-poor star with a mass of $\sim0.01~M_\odot$. 

\begin{figure}[ht]
    \centering
    \includegraphics[width=0.99\columnwidth]{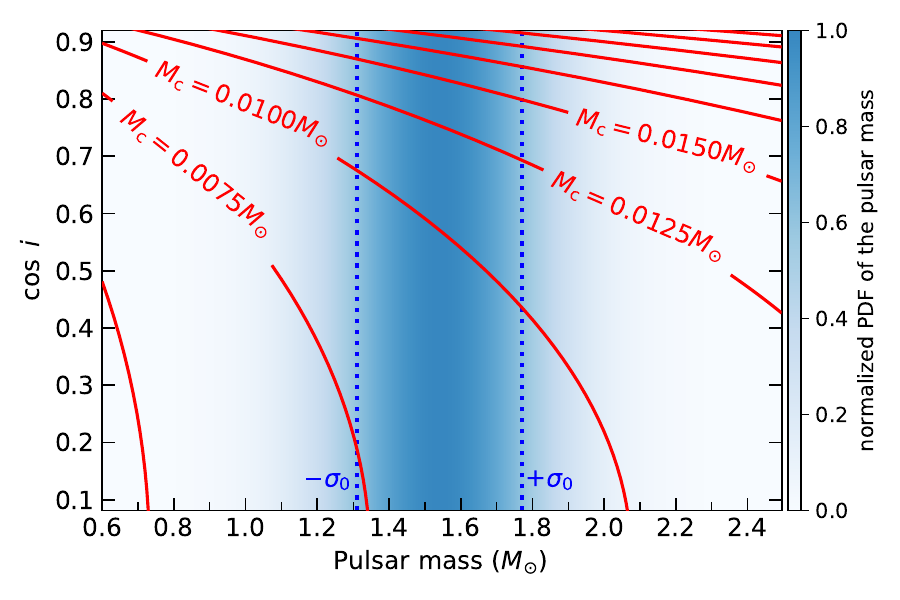}
    \includegraphics[width=0.85\columnwidth]{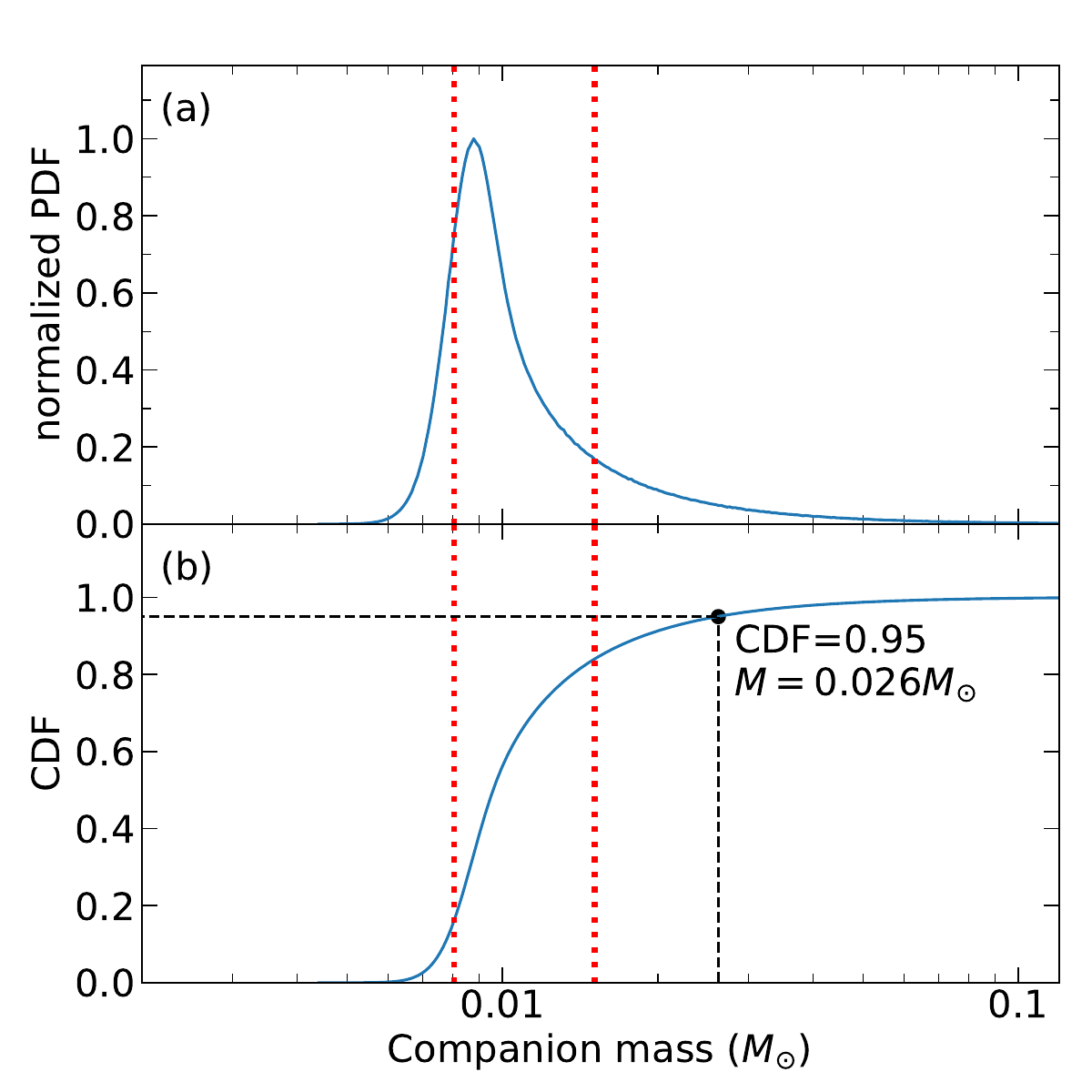}
    \caption{ {The probability distribution of the companion mass $M_{\rm c}$ of M71E.} {\it Top panel:} distribution of possible companion mass calculated for various inclination angle $i$ and pulsar mass $M_{\rm psr}$. Here, $\cos i$ should follow a uniform probability distribution if the orbital orientation is random in space. The probability density function (pdf) for the pulsar mass presented in the gray scale is taken from \citet{of16} for the recycled pulsars, following a Gaussian function centering at $M_0 = 1.54~M_{\odot}$ with a standard deviation of $\pm\sigma_0 =0.23~M_{\odot}$ as indicated by two dotted lines.
    {\it Bottom panel:} the PDF (in subpanel (a))  and the cumulative distribution function (CDF, in subpanel (b)) for the companion mass $M_{c}$ integrated from the pdf on the top panel, with a probability of $\cos i$ uniformly distributed in the range of (0, 1). Two red dotted lines indicate the companion mass of  $0.008$ and $0.015~M_{\odot}$ at the 16\% and 84\% probability ($\pm1\sigma$) of the CDF and the dashed line marked the 95\% probability of the CDF at 0.026 $M_\odot$. 
    }
    \label{mc}
\end{figure}

\begin{figure}[ht]
    \centering
    \includegraphics[width=0.99\columnwidth]{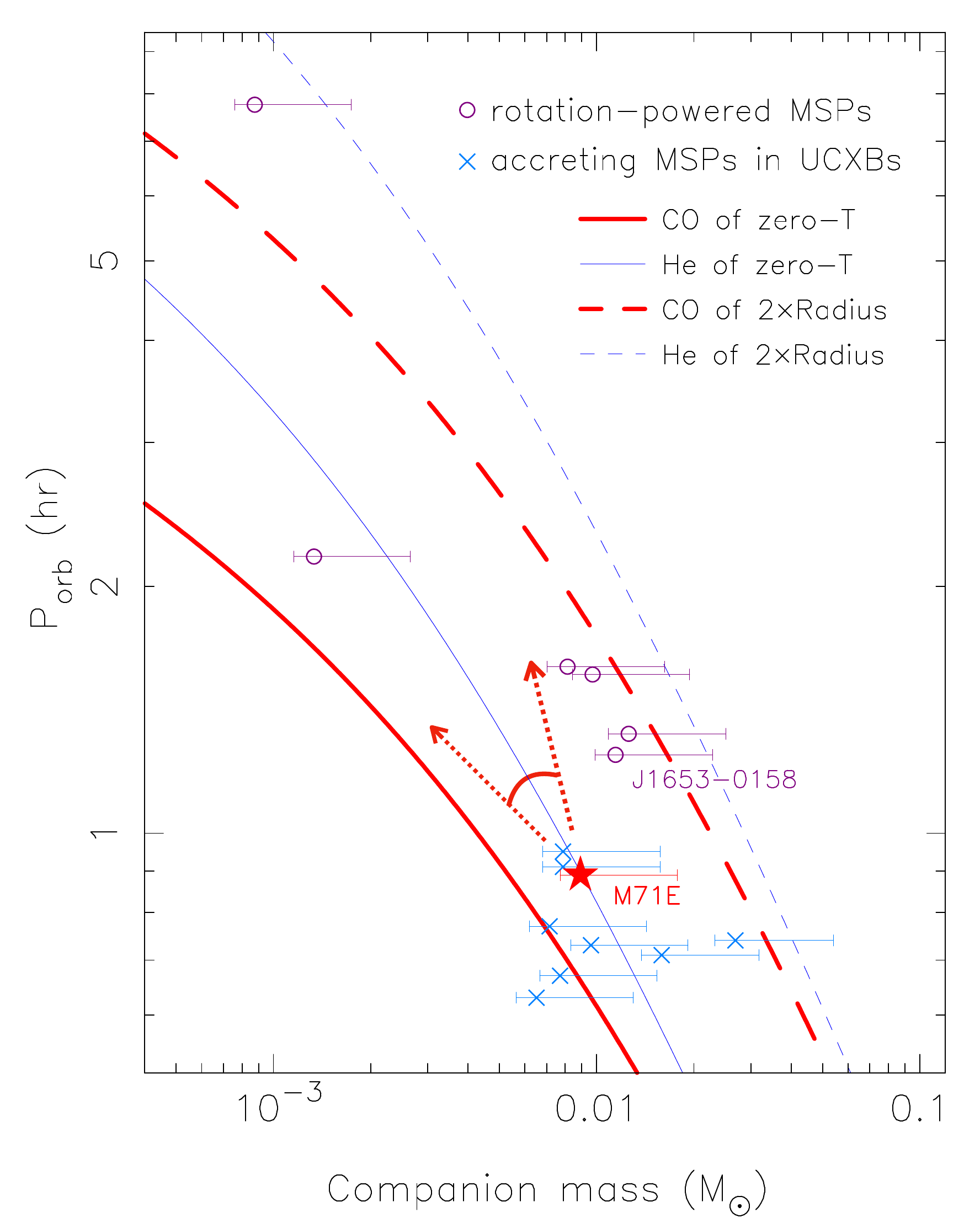}  
    \caption{ {Companion mass vs. orbital period for accreting MSPs in UCXBs and possible rotation-powered pulsar descendants.} Data are taken from \cite{2022MNRAS.515.2725G}, ATNF pulsar catalog \citep{2005AJ....129.1993M}, and also Table 4.1 in  \citet{2022ASSL..465.....B}. The error bars represent the mass range of these systems from $\cos i=0$ to $\cos i=0.9$ (90\% possibility) assuming a 1.4 $M_\odot$ neutron star mass, with the median companion mass calculated at $\cos i=0.5$  marked by circles and crosses for rotation-powered MSPs or accretion-powered MSPs, respectively. The binary system with a zero-temperature (zero-T) companion with a filled Roche lobe is indicated by solid curves, while that holds a companion with a zero-T radius of only one-half of the Roche lobe radius is indicated by dashed curves \citep{1971ARA&A...9..183P,1987ApJ...322..842R}. M71E is marked as a star. The  {parameter} range roughly indicated by the two arrows is the possible long-term evolution direction in the parameter space.}
    \label{pb-m2}
\end{figure}

\section{FAST results and binary constraints}

M71E has a spin period of 4.44 ms \citep{2021RAA....21..107H}. It has been suspected to be not physically associated with the globular cluster  \citep{pan21} though its dispersion measure (DM) is very close to the DMs of known pulsars inside the cluster. More follow-up observations by the FAST over 1.5 yr yield an excellent phase-link timing solution, including the derivative of pulsar spin period  ($1.1\times10^{-15} {\rm~s}^{-2}$) and exact position. At the outskirt of M71 at a distance of 4 kpc, the kinetics contribution on the spin frequency derivative from the global cluster is at most $2\times10^{-16} {\rm~s}^{-2}$ \citep{2018MNRAS.478.1520B}, roughly 1 order smaller than the measured value. We therefore trust that a spin-down luminosity of  $\dot E = I\Omega\dot \Omega=1\times10^{34}~{\rm erg~s}^{-1}$ is reliable,  {a typical value} for spider pulsars.

The mass function from the timing solution of M71E is $2.3\times10^{-7}~M_\odot$ \citep{53-min}, which is related to the companion mass, pulsar mass, and orbital inclination by
\begin{equation}
    f(m)=(M_{\rm c}\sin i)^3/(M_{\rm NS}+M_{\rm c})^2.
    \label{mf}
\end{equation}
Using the recycled-pulsar mass distribution from \citet{of16} and a random orbital orientation, we get that its companion mass is in the range of $8-15\times10^{-3}~M_\odot$ at the 68\% confidence level ($\pm1\sigma$) and \textless$0.026~M_\odot$ at the 95\% confidence level, as shown in Fig.~\ref{mc}. The short orbital period of M71E and a flyweight companion of mass $\sim0.01~M_\odot$ bear a striking resemblance to many UCXBs (see Fig.~\ref{pb-m2}). 

At the accurate timing position of pulsar, a faint X-ray source \citep{53-min} and an optical counterpart \citep{liu2023optical} are found. The X-ray counterpart that we identified for this system is 2CXO J195337.9+184454 from \citet{2010ApJS..189...37E}, which has a X-ray luminosity of $1\times10^{30} \rm~erg~s^{-1}$ in 0.5$-$7.0 keV, about $10^{-4}$ of M71E's spin-down luminosity.  {We noticed that \citet{2023arXiv230913189Z} also identified this source. They fitted a blackbody-like model and got a X-ray luminosity of $1.9^{+1.2}_{-1.1}\times10^{30} \rm~erg~s^{-1}$, similar to those of other noneclipsing black widows \citep{2023arXiv230913189Z}.}

The optical counterpart  {identified by \citet{liu2023optical}} should have a variation if the orbit is not face on, though it has not yet detected. The temperature difference between the heated side (dayside) and the unheated side (nightside) can be estimated base on the spin-down power $\dot E$ of the pulsar \citep{2016ApJ...828....7R},
\begin{equation}   
    T^4_{\rm D} - T^4_{\rm N} =\eta\dot E/4\pi a^2\sigma_{\rm SB},
    \label{T_d}
\end{equation}
where $T_{\rm D}$ and $T_{\rm N}$ are the effective temperatures of the heated side and unheated side respectively, $\eta$ is the heating efficiency, and $a$ is the orbital separation. The heating efficiency varies from system to system, and in some systems it can be even larger than 1 due to anisotropic emission \citep{2016ApJ...828....7R}. Assuming a Roche lobe filling factor of $f$, a pulsar mass of 1.4 $M_\odot$ and a companion mass of 0.01 $M_\odot$, we find that the heated side temperature should be about $\rm10^4~K$ for a heating efficiency about 1, and the equivalent luminosity of the companion star on the heated side is at least $0.02\eta f^2~L_\odot$, which is very sufficient to explain the luminosity of its optical counterpart with an assumption of blackbody-like radiation with $E_{10^4k}/E_{4500k}(\lambda=606nm)=0.82~L_{10^4K}/L_{4500K}$. \citep{liu2023optical}.

\section{Evolution of M71E}

Referring to Fig.2, the M71E binary system has an orbital period even shorter than some known UCXBs, which makes the formation and evolution picture more complicated. This radio MSP probably was formed from a UCXB at an earlier evolutionary stage in the transition from AMXPs to radio MSPs. 

\subsection{The Mass Transfer Phase}

Depending on different types of donor stars, such as white dwarfs (WDs) or helium stars or main-sequence (MS) stars \citep{1975MNRAS.172..493P,1986ApJ...304..231N,1986A&A...155...51S,2002ApJ...565.1107P,2002A&A...388..546Y,2008AstL...34..620Y,2017MNRAS.470L...6S,2022ApJ...930..134C,2022MNRAS.515.2725G}, the evolution routines of UCXBs are quite different. WD donors become fully degenerate very quickly. However, with the helium star donor and the MS donor, the system should experience a long time of mass transfer from the donor star, and then is degenerated and thermally relaxed. Afterwards, the evolution tracks for various UCXBs in these scenarios are very similar \citep{2012A&A...537A.104V}. In a stable mass transfer phase, the donor star has a radius $R_{\rm c}$ roughly equal to its Roche lobe radius $R_{\rm L}$. Given the well-known mass-radius relationship for the degenerated star $R\propto M^\beta~(\beta=\rm-1/3)$, the donor stars should expand in response to the mass loss \citep{2003ApJ...598.1217D,2012A&A...537A.104V,2013ApJ...768..184H}. As a result, their orbital periods increase to keep $R_{\rm c}\approx R_{\rm L}$. 

\subsection{The Roche Lobe Detachment Phase}

With a circular orbit, the orbital angular momentum of a UCXB for a neutron star (NS) accretor can be written as 
\begin{equation}
    J^2_{\rm orb}=\frac{GM_{\rm c}^2M^2_{\rm Nd}}{M_{\rm tot}}a,
    \label{J_orb}
\end{equation}
where $J_{\rm orb}$ is the orbital angular momentum, $a$ is the orbital separation, $M_{\rm c}$ is the companion mass, and $M_{\rm Nd}$ is the sum of the NS mass $M_{\rm NS}$ and the accretion disk mass $M_{\rm disk}$ in the vicinity of the NS. The combination of this equation and the Kepler's third law leads to
\begin{equation}
    2\frac{\dot J_{\rm orb}}{J_{\rm orb}}=2(\frac{\dot M_{\rm c}}{M_{\rm c}}+\frac{\dot M_{\rm Nd}}{M_{\rm Nd}})-\frac{\dot M_{\rm tot}}{M_{\rm tot}}+\frac{\dot a}{a}
    \label{evo}
\end{equation}
and
\begin{equation}   
    3\frac{\dot a}{a}-2\frac{\dot P_{\rm orb}}{P_{\rm orb}}=\frac{\dot M_{\rm tot}}{M_{\rm tot}}.
    \label{evo_a}
\end{equation}

In the detachment phase, we take $R_{\rm c}=R_{\rm L}$ and $\dot R_{\rm c}<\dot R_{\rm L}$. Given that $R_{\rm c}\propto M_{\rm c}^\beta$ and in the case $M_{\rm c}\ll M_{\rm Nd}$, one gets $R_{\rm L}\propto a(M_{\rm c}/M_{\rm Nd})^{1/3}$ \citep{1971ARA&A...9..183P,1983ApJ...268..368E}. That is to say 
\begin{equation}
    \beta\frac{\dot M_{\rm c}}{M_{\rm c}}<\frac{1}{3}(\frac{\dot M_{\rm c}}{M_{\rm c}}-\frac{\dot M_{\rm Nd}}{M_{\rm Nd}})+\frac{\dot a}{a}.
\end{equation}

With a flyweight donor, the orbital angular momentum loss is dominated by mass loss. Similar to \cite{1994inbi.conf..263V} and \cite{1997A&A...327..620S}, one can get 
\begin{equation}
    \delta J_{\rm orb}= \delta M_1\frac{M_{\rm Nd}}{M_{\rm c} M_{\rm tot}}J_{\rm orb}+\delta M_2\frac{M_{\rm c}}{M_{\rm Nd}M_{\rm tot}}J_{\rm orb},
    \label{D-J}
\end{equation}
where $\delta M_1$ is the mass lost from the donor star as fast isotropic winds, and $\delta M_2$ is the mass ejected from the vicinity of the NS in an isotropic way. At the time of detachment, there is no Roche lobe overflow, therefore the main mass loss is the evaporation of the companion caused by the pulsar wind. This  leads to $\delta M_1=\delta M_{\rm c}$ and $\delta M_2=\delta M_{\rm Nd}=0$. Combining them with Eq.~\ref{evo}, one can obtain ${\dot a}/{a}=-\dot M_{\rm c}/{M_{\rm tot}}$. Then in Eq.~\ref{D-J}, one gets $\beta>\frac{1}{3}$. 

For degenerated stars the mass-radius relationship requires $\beta<0$ until the mass of object is below the maximum-radius mass value \citep{1987ApJ...322..842R}, which should not be the case for the companion mass of M71E. The donor star should have a nondegenerate outer layer so that the detachment scenario becomes possible. The shrinkage of the companion radius may be not driven by mass loss, but could be a decrease in the heating by the pulsar or a more efficient cooling \citep{2012A&A...541A..22V}. Such a bloated companion is also supported by a high temperature derived from Eq.~\ref{T_d} and optical observations of similar systems \citep{2023ApJ...942....6K}, and this system may experience cycles of expansion and shrinkage as suggested by \citet{2000A&A...360..969R}. In Fig.~\ref{pb-m2}, an bloated companion should be in the very right side of the zero-temperature (zero-T) line in Fig.~\ref{pb-m2}. However, M71E is located by chance on the He zero-T line in Fig.~\ref{pb-m2}, and in the very right side of the CO zero-T line; therefore a CO WD is more likely for any inclination angles. If it is a He WD, a very low inclination angle is strongly desired, which is much less likely to exist. 

Another possible mechanism for the detachment of M71E's companion involves the disk instability \citep{2012A&A...541A..22V,2022MNRAS.515.2725G}.  A stable helium-composition disk requires the lowest mass transfer rate \citep{2001NewAR..45..449L,2002ApJ...564L..81M,2007A&A...465..953I,2012A&A...537A.104V}
\begin{equation}
    \dot M_{\rm crit}\approx3\times10^{-10}(\frac{M_{\rm c}}{M_\odot})^{0.3}(\frac{P_{\rm orb}}{\rm hr})^{1.4}~M_\odot~{\rm yr}^{-1}, 
\end{equation}
below which a thermal-viscous instability appears and the disk experiences outbursts. The outburst causes a sudden decrease of the disk mass; therefore $\delta M_1=\delta M_{\rm c}=0$ and $\delta M_2=\delta M_{\rm Nd}\geq-M_{\rm disk}$. The material is ejected from the vicinity of the NS, taking away its orbital angular momentum \citep{2021MNRAS.506.4654W}. Again using Eq.~\ref{evo}, we obtain $\delta a=a(-\delta M_2/M_{\rm tot})$, which leads to a sudden increase of the Roche lobe radius without changes in companion size. The companion becomes detached.
This detachment scenario may lead to a similar behavior of transitional MSPs \citep{2009Sci...324.1411A} which switch back and forth between X-ray and radio pulsar state \citep{2012A&A...541A..22V}. 

\subsection{Roche Lobe Reattachment}

The companion radius should be in the range between its zero-temperature radius and Roche lobe radius, providing a constraint of its density $\rho_{{\rm c,zero}{\text -}T}>\rho_{\rm c}>\rho_{{\rm c},R_{\rm c}=R_{\rm L}}$. 

After the Roche lobe detachment, the long-term evolution is dominated by the evaporation of the companion caused by the pulsar wind. The companion mass is lost by fast isotropic winds from the companion, and the lost mass also carries away the orbital angular momentum; using Eq.~\ref{evo}, \ref{evo_a}, and \ref{D-J}, we obtain ${\dot P_{\rm orb}}/{P_{\rm orb}}=-2\dot M_{\rm c}/{M_{\rm tot}}$, the same as \citet{1924MNRAS..85....2J}. This has been widely taken as the contribution from the companion mass loss on the variations of the orbital period in spider pulsar systems \citep{2011MNRAS.414.3134L,2015ApJ...807...18P,2023ApJ...942...87K}. When this mass-loss effect dominates the variations of orbital period, one gets $P_{\rm orb}\propto M_{tot}^{-2}$. The changes in total mass of this system are relatively small, so the orbital period nearly freezes while the companion mass continues to decrease. Before the system crosses the solid curves (reaching $\rho_{{\rm c,zero}{\text -}T}\leq\rho_{{\rm c},R_{\rm c}=R_{\rm L}}$), a reattachment of M71E's companion is inevitable. If the companion star has a size close to its Roche lobe, the Roche lobe reattachment may not be driven by mass loss and may happen much more quickly as a part of cyclic behaviors \citep{2000A&A...360..969R,2012A&A...541A..22V}. After that, a new Roche lobe overflow phase begins, during which its orbital period will increase to keep $R_{\rm c}\approx R_{\rm L}$ as discussed in Sec.~3.1. As shown roughly by the two red arrows in Fig.2, the system will evolve toward the upper left corner in this parameter space, and becomes a system similar to other descendants of UCXBs, such as PSR J1653–0158 \citep{2020ApJ...902L..46N}.

\section{Conclusion}

Discovered during the FAST GPPS survey \citep{2021RAA....21..107H}, M71E is a remarkable binary pulsar due to the shortest orbital period among all known rotation-powered pulsar binaries \citep{53-min}. 
According to the mass function, the companion mass of M71E is $\sim0.01~M_\odot$, and the orbit of this system is strikingly similar to that of AMXPs in ultra-compact X-ray binaries (see Fig.\ref{pb-m2}). Therefore a probable alternative explanation is that M71E is a descendant from a UCXB. This was briefly mentioned in \citet{53-min} and now discussed in detail here as a complementary theory.

In this scenario, M71E has a very short orbital period so it is in a much earlier evolutionary stage compared to other radio MSPs that are descendent of UCXBs. The possible evolution of M71E is that a natural detachment from the Roche lobe happens when $\beta>\frac{1}{3}$, which means that the companion density increases with the mass loss. A nondegenerate outer layer is required in this scenario, and it favors a CO-rich companion. If it is a He WD, a low inclination angle of the orbit is strongly desired, which is much less likely to exist. Another possibility involves X-ray outbursts caused by the disk instability, and we then expect M71E to behave like a transitional MSP. In both cases, a reattachment of M71E's companion is inevitable in the future. We suggest that in the late evolution stage, the MSPs in UCXBs will experience several transitions between a radio pulsar and an accretion-powered X-ray pulsar.

\section*{Acknowledgements}
The authors are supported by the Natural Science Foundation of China, Nos. 11988101, 11833009, and also the National SKA Program of China, No. 2022SKA0120103. 
We thank Dr. YunLang Guo and Prof. HaiLiang Chen for helpful comments.

\end{document}